\documentclass[reprint,amsmath,amssymb,aip,pre,author-numerical]{revtex4-1}
\pdfoutput=1

\usepackage{epsfig}
\usepackage{hyperref}
\usepackage{amsfonts}
\usepackage{amsmath}
\usepackage{color}
\usepackage{mathtools}
\usepackage{siunitx}
\usepackage{dcolumn}
\usepackage{graphicx}
\usepackage{float}
\usepackage{bm}
\usepackage[dvipsnames]{xcolor}

\usepackage{verbatim}
\immediate\write18{texcount -tex -sum  main.tex > \jobname.wordcount.tex}

\begin{document}
	
	\title{Nonlocal dielectric properties of water: the role of electronic delocalisation}

\author{Darka Labavic}
\affiliation{UMR CNRS Gulliver 7083, ESPCI Paris, PSL Research University, 75005 Paris, France}%
\author{Florian N. Brünig}
\affiliation{Fachbereich Physik, Freie Universität Berlin, Arnimallee 14, Berlin, 14195, Germany}
\author{Roland R. Netz}
\affiliation{Fachbereich Physik, Freie Universität Berlin, Arnimallee 14, Berlin, 14195, Germany}
\author{Marie-Laure Bocquet}
\email{marie-laure.bocquet@phys.ens.fr}
\affiliation{Laboratoire de physique de l'école normale supérieure, ENS, Université PSL, CNRS, Sorbonne Université, Université Paris Cité, Paris, France}%
\author{Hélène Berthoumieux}
\email{helene.berthoumieux@espci.psl.eu}
\affiliation{UMR CNRS Gulliver 7083, ESPCI Paris, PSL Research University, 75005 Paris, France}%

\begin{abstract}
	The nonlocal dielectric properties of liquid water are studied in the context of {\it ab initio} molecular dynamics simulations based on density functional theory. We calculate the dielectric response from the charge structure factor of the liquid using the fluctuation-dissipation theorem. We show that the dielectric response function of {\it ab initio} simulations differs significantly from that of classical force-fields, both qualitatively and quantitatively. In particular, it exhibits a larger amplitude and a wider range of responding wave numbers. We suggest that the difference is due to the localisation of the electronic charge density inherent in classical force files and Wannier post-treatment of DFT densities. The localised charge models do not reproduce the shape of the response function even for $q$ corresponding to intermolecular distances, and could lead to a significant underestimation of the dielectric response of the liquid by a factor of 10.
\end{abstract}	

	\maketitle
		
\section{Introduction}
Characterizing the screening properties of water at microscopic scales is essential for understanding many interfacial  dynamical phenomena at electrode surface like friction and its recently unveiled quantum counterpart,~\cite{kavoline2022} like redox reactions entailing electron transfer.~\cite{phelps1990,jeanmairet2018} A key observable to account for the water electrostatic contribution is the dielectric response function of bulk water from {\it ab initio} molecular dynamics (AIMD) simulations that include the electronic density.  \par
The static dielectric constant of water, $\epsilon$, is well-known about 80 for macroscopic distances, {\it i.e.} the bulk regime. However, at nanometric range, the screening properties of the fluid deviates dramatically from the bulk environment.~\cite{kornyshev1981,vatin21} A convenient way of describing this scale-dependent behavior is to introduce a non-local permittivity  $\epsilon(q)$ for liquid water  which is a function of the wavenumber $q$ in Fourier space. 
 The shape of $\epsilon(q)$ governs ion-water and ion-ion interaction at the nanoscale but also the structure of interfacial and confined solutions.~\cite{schlaich2016,fumagalli2018,monetprl,hedley2023,robert23}\par
\begin{figure*}
	\center
\includegraphics{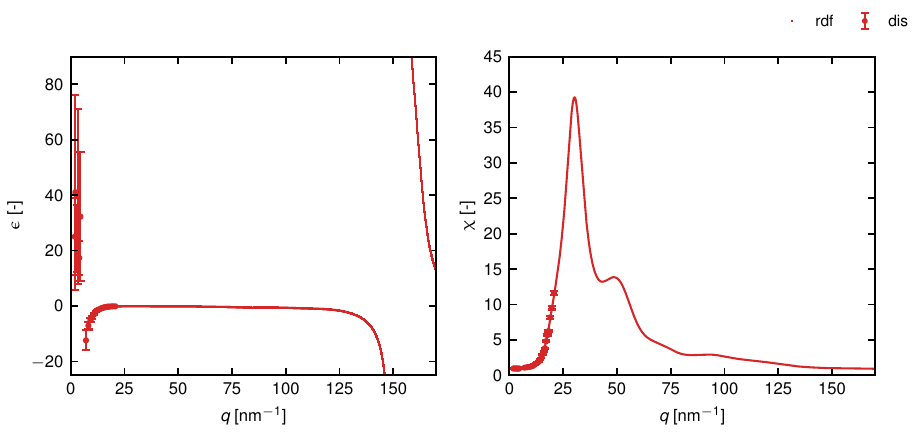}
\caption{Dielectric permittivity $\epsilon(q)$ (a) and dielectric response function $\chi(q)$
	(b) of bulk water as a function of the wavenumber $q$ from classical force-field SPC/E simulations
	of water. Simulation methods are given in ref~\cite{hedley2023}. $\epsilon(q)$ and $\chi(q)$ are calculated from the radial distribution function for large q (solid line) and for discrete wavenumbers at low q (data points).}
\label{fig1:eps_chiSPCE}
\end{figure*}

Using the fluctuation-dissipation theorem, $\epsilon(q)$ can be expressed as a function of the charge structure factor $S(q)$, which is easily computed from classical molecular dynamics (MD) simulations. $\epsilon(q)$ for water, firstly simulated in 1996 using the classical force-field BJH~\cite{bopp1996} reveals  a range of wavenumbers $q$ over which the permittivity  $\epsilon(q)$ is negative, which was referred to as an "overscreening zone". As an illustration, Fig.~\ref{fig1:eps_chiSPCE} a) represents $\epsilon(q)$ for the SPC/E model of water and exhibits a range of negative screening values, and is flanked by diverging poles at low and high $q$.

The dielectric response function $\chi(q)$, also called the dielectric susceptibility, is free of poles ~\cite{bopp1996,kornyshevtrans} and is a probe of water charge structure at the molecular level. It depends directly on the charge structure factor as $\chi(q)=S(q)/q^2\epsilon_0k_BT$ and is related to $\epsilon(q)$ by $\chi(q)=(1-\epsilon(q))^{-1}$. 
$\chi(q)$ for SPC/E water model is plotted in Fig.~\ref{fig1:eps_chiSPCE} b). We observe a zone of "overresponse" ($\chi \gg1$) corresponding to the overscreening range for $\epsilon(q)$. $\chi(q)$ has a continuous resonant form that reaches a sharp maximum for $q$=$\SI{30}{nm}^{-1}$, indicating a typical length $\lambda = 2\pi / q \simeq$\SI{20}{nm} for the charge organisation: the H-bond length (indeed the distance O-O in O-H...O hydrogen bound ranges from 3 to 3.5 Å).

 Both $\epsilon(q)$ and $\chi(q)$ observables have been computed via MD atomistic simulations of bulk water for rigid,~\cite{monetprl,becker} flexible~\cite{bopp1996} and polarizable~\cite{elton_these}  classical force-field models, showing consistently a phenomenon of "overscreening"/"oversponse".
Up to now, the effect of the delocalised electronic charge on $\chi(q)$ is not included in such models although it is expected to be important, in particular at high wavenumbers,~\cite{kornyshev1999} i.e. at sub-angstrom distances.

The study of the dielectric properties of water using AIMD simulations initially targets macroscopic and static properties of liquid water such as the molecular dipole~\cite{silvestrelli1999} and the dielectric constant.~\cite{sharma2007} But in recent years, AIMD simulations have been applied to investigate nanoconfined water within graphene slits.~\cite {barragan2020,dufils2024,BeckerJCP2024}
For the analysis of the charge distributions in AIMD simulations, the electronic density of water is usually treated using the maximally localized Wannier function formalism. This post-processing treatment assigns the 8 valence electrons of water molecule to localized chemical bonds: 2 electrons for each OH bonds and 2 electrons per oxygen lone pair.~\cite{sharma2003} In most cases, the resulting Wannier dipole moment (computed from the charges Wannier centers and the nuclei) is the only observable output which is sufficient for getting the total $\epsilon(q \to 0)$ that is compared to measurements. \par
It is well-studied  that the dielectric properties of water calculated with {\it ab initio} simulations show important differences when compared to the classical MD simulations, both for bulk~\cite{carlson2021} and confined water,~\cite{dufils2024} for nonlocal and for frequency-dependent properties.~\cite{carlson2021,dufils2024} Moreover, AIMD simulations better agree with experimental data than the classical force-fields.~\cite{carlson2021,dufils2024} This is due in particular to a better representation of hydrogen bonding, which leads to an improved description  of the collective effects of the H-bond network. All these charge correlation effects are encoded in the dielectric response function $\chi(q)$, which has not yet derived from AIMD.\par
 
In this work, we compare bulk liquid water's nonlocal dielectric response function for SPC/E classical force-field and for quantum DFT simulations.
The results between AIMD and classical MD simulations differ significantly, quantitatively and qualitatively.  To understand this unexpected result, we compute the dielectric response for different classical force fields to tackle the effect of the charge distribution inside a water molecule.
For sake of comparison, we renormalise the $\chi(q)$ spectra of the different models by the total electronic charge per molecule and obtain a "master" response function plot. We use this plot to discuss the effects of structural flexibility and electronic delocalisation included in the  DFT-based AIMD method that are missing in rigid classical models. We show that the susceptibility $\chi(q)$ associated with the Wannier center yields spurious results, due to non-physical high values of localized electronic charges. On the contrary, the response function obtained from the Wannier-derived dipoles alone can be compared very well with the SPC/E spectrum. Finally, we show that more advanced classical force fields as flexible or polarisable models do not describe the electronic correlations significantly better than SPC/E. Taken together, this work shows that AIMD with direct electronic density is the only way to get the correct dielectric response of water, in the width and amplitude of the response function.
 \par

\section{Simulation methods}
\begin{figure*}
	\center
	\includegraphics[width=0.25\textwidth]{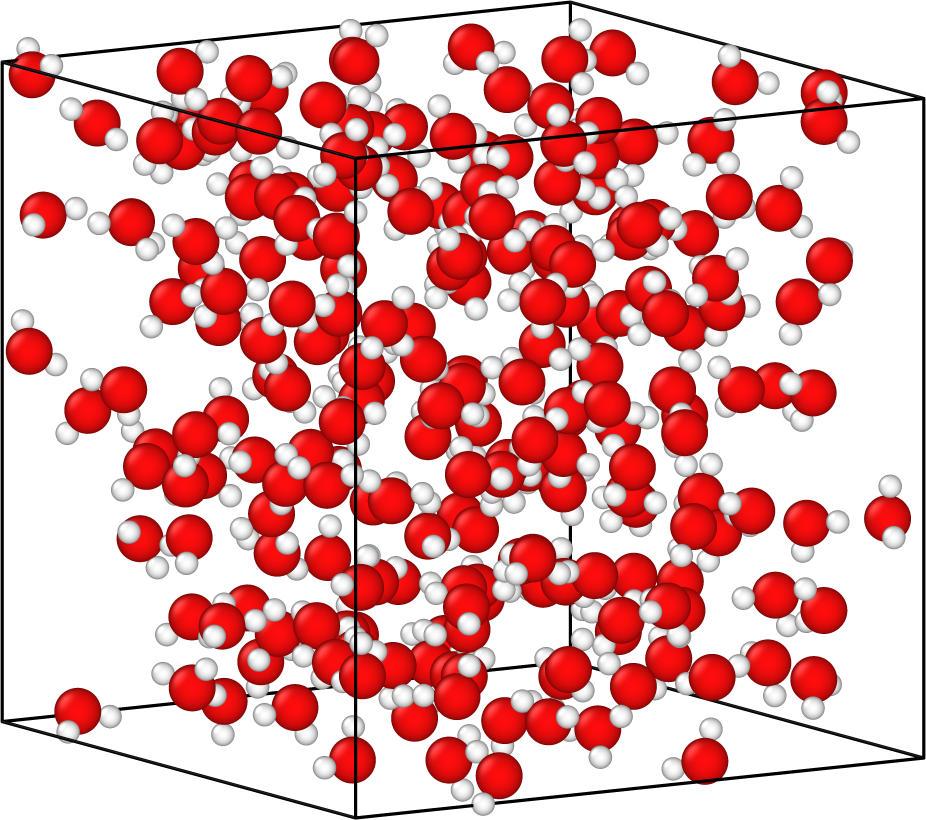}
	\hspace*{1cm}
	\includegraphics[width=0.25\textwidth]{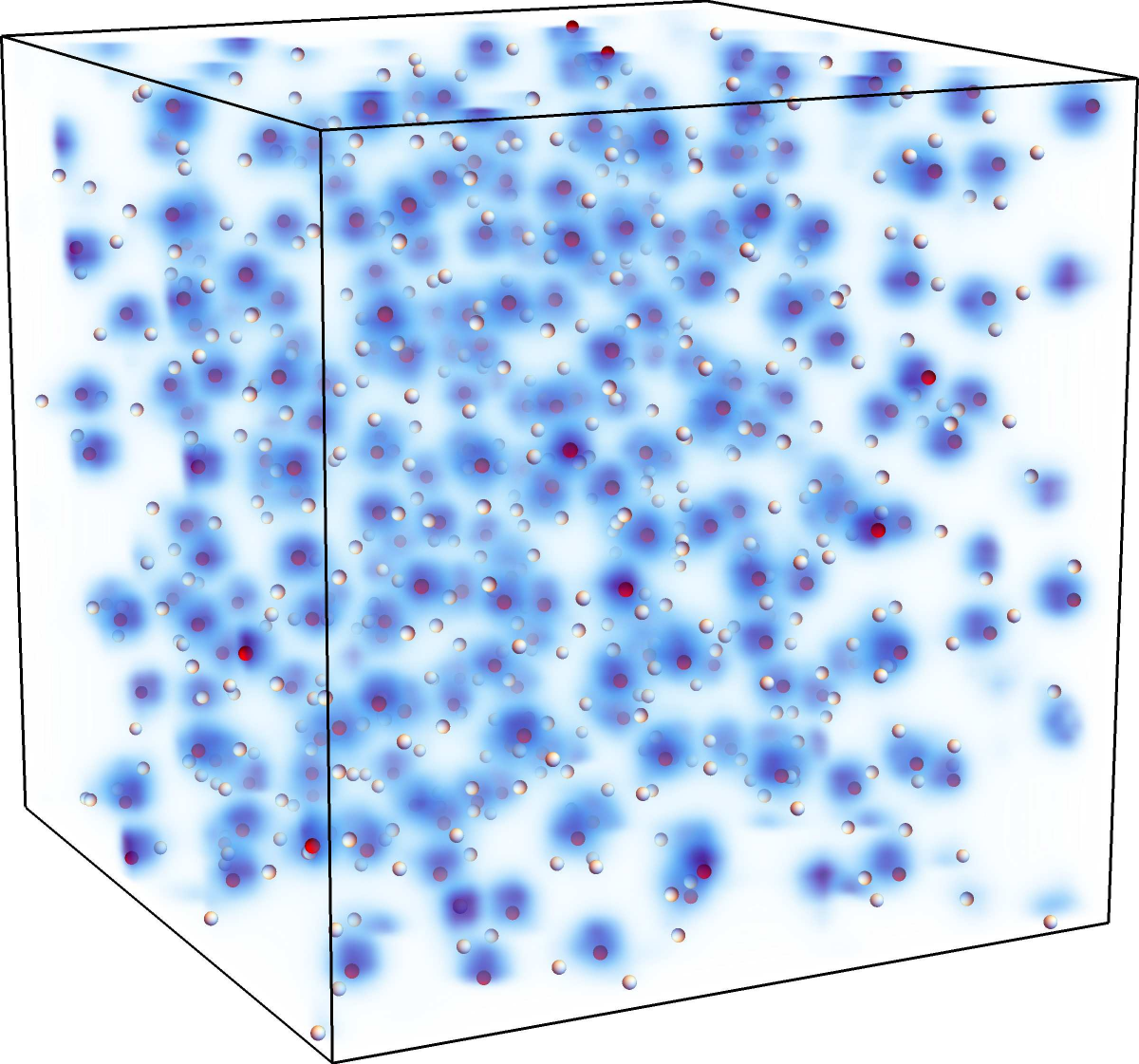}
	\hspace*{1cm}
	\includegraphics[width=0.2\textwidth]{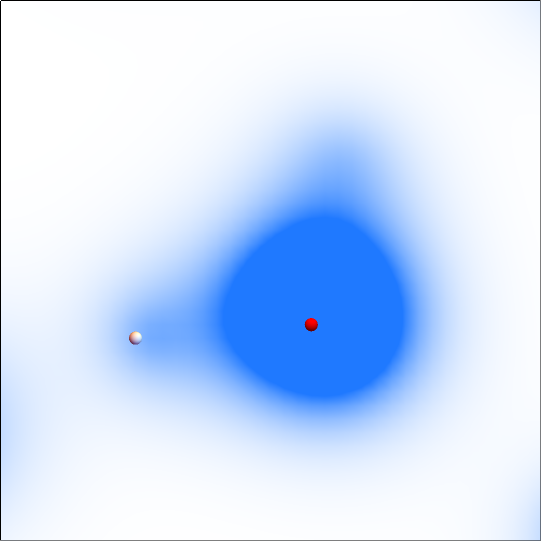}
	\caption{Snapshot of the classic and quantum simulation boxes of bulk water. Size of the box: $L$=2~nm. (a) SPC/E water. Red and white balls represent the water molecules. (b) Charge density for an AIMD simulation box. (c) Zoom on a single molecule from (b). The red and white points represent the nuclei of Oxygen and Hydrogen. The blue cloud shows the electronic density. }
		\label{simulatedsystem}
\end{figure*}

\begin{figure}
	\center
	\Large{Classical models} \\
	\includegraphics[width=.15\textwidth]{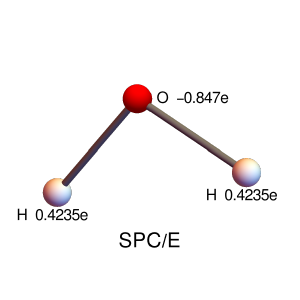}
	\includegraphics[width=.15\textwidth]{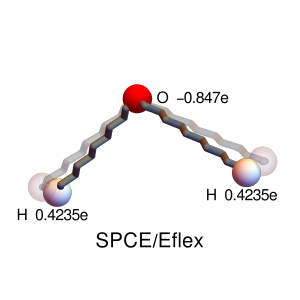}\\
	\includegraphics[width=.15\textwidth]{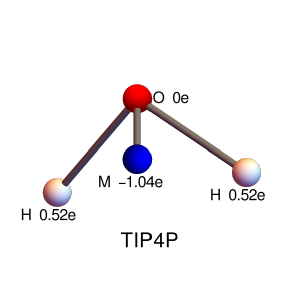}
	\includegraphics[width=.15\textwidth]{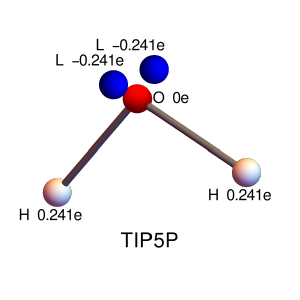}\\
	\Large{DFT} \\
	\includegraphics[width=.15\textwidth]{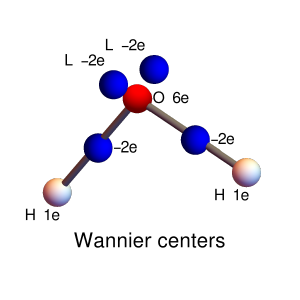}
	\includegraphics[width=.20\textwidth]{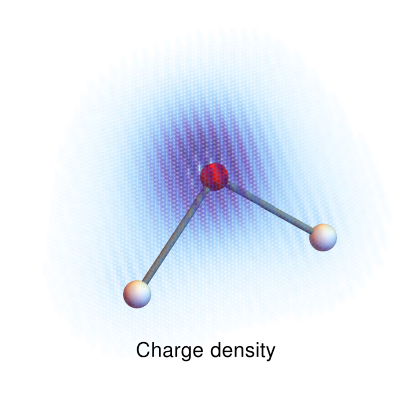}\\
	\caption{Water models for each water molecule in bulk liquid. This a sketch of the 4 classical force field models and the 2 {\it ab initio} charge distributions considered here. For SPC/E, SPC/E flex, TIP4P, TIP5P, we give the partial charges associated with the Oxygen O (red sphere), with the hydrogens H (white spheres), and with the Lone pairs and the Dummy site (blue sphere). For the wannier projection, there is 4 Wannier centers (blue sphere) on a particular water molecule : two correspond to OH bonds and other two can be identified as lone pair electrons in the oxygen atom. Hence water is modeled as 7 point charges : 4 wannier functions and 3 charges on each atoms.}
	\label{water_molecule}
\end{figure}

\subsection{Classical force-field MD simulations}
We simulate a cubic water box of size $L$=\SI{2}{nm}. See a snapshot of the simulated system in Fig.~\ref{simulatedsystem} described below and pictured in Fig.~\ref{water_molecule}.  We perform MD simulations with classical force-field water models~\cite{loche2021}. Simulations are performed using the GROMACS 2023 molecular dynamics simulation package~\cite{gromacs}, and the integration time step is set to $\Delta t$=\SI{2}{fs}. All simulation boxes are periodically replicated in 3 directions, and long-range electrostatics are handled using the smooth particle mesh Ewald (SPME) technique. Lennard-Jones interactions are cut off at a distance $r_{\rm cut}$=\SI{0.9}{nm}.  A potential shift is used at the cut-off distance. All systems are coupled to a heat bath at 300 K using a v-rescale thermostat with a time constant of 0.5 ps. We use MDAnalysis to treat the trajectories. After creating the simulation box, we perform a first energy minimization. Next, we equilibrate the system in the NVT ensemble for 200 ps and afterward in the NPT ensemble for another \SI{200}{ps} using a Berendsen barostat at 1 bar. Production runs are performed in the NVT ensemble.\par
We consider the SPC/E water model as the reference one for classical force-field simulations, see sketch in Fig.~\ref{water_molecule}. We chose SPC/E over more complex models because of its simplicity, its good prediction of the dielectric constant, and its popularity in classical MD simulations.
SPC/E contains 3 point charges, one on the oxygen and one on each hydrogen. The Lennard-Jones (LJ) center is placed on the oxygen.
\par
We also consider a flexible version of the SPC/E model which we denote as SPC/Ef. In this modification, the OH bond length and the $\hat{HOH}$ angle can vary and are described by the bond and angle harmonic potentials.
\begin{align}
\label{k_flex}
V_d=\frac{k_d}{2}(d-d_0)^2, \quad V_\theta=\frac{k_\theta}{2}(\theta-\theta_0)^2,
\end{align}
with the bond streching force constant $k_{\rm OH}$=$\SI{345000}{kJ\per mol \per \nm^2}$ and an angle-bending force constant $k_{\rm HOH}$=$\SI{417.6}{kJ\per mol \per rad ^2}$\par


Second, we consider another 3-point-charge model associated with slightly different parameters, TIP4P. It is a 4-site water model as shown in Fig.~\ref{water_molecule}. The LJ center is placed on the oxygen. Positive charges are placed on the hydrogen atoms, and an additional interaction site, M, and carries the negative charge.\par
Finally, TIP5P is non-planar 5-interaction site, including 4 charges, was developed to reproduce the density anomaly for water~\cite{mahoney2000}. A LJ center is localized on the oxygen. Positive charges are placed on the hydrogen atoms and negative charges on additional sites, L, representing the lone pairs. \par

Note that the number N$_w$ of water molecules in the simulation box varies depending on the model (221 for SPC/E, SPC/E flexible and TIP4P, 256 for TIP5P)

Two simulation runs of varying durations are carried out for the classical force-fields models: a short run of \SI{200}{ps} and a long run of \SI{20}{ns}. In both cases, we use a time step of \SI{2}{fs} and a write-out frequency of \SI{2}{ps}.
A third simulation is carried out for the flexible model: a short run of \SI{200}{ps} but with a time step of \SI{0.5}{fs} and a write-out frequency of \SI{2}{fs}.

\subsection{DFT MD simulations}
We simulate a cubic water box of size $L$=\SI{2}{nm} composed of 256 water molecules. The DFT MD simulation is carried out in CP2K 4.1 using a polarizable double-$\zeta$ basis set for the valence electrons, optimized for small molecules and short ranges (DZVP-MOLOPT-SR-GTH), dual-space pseudopotentials, the BLYP exchange-correlation functional, and D3 dispersion correction.  The cutoff for the plane-wave representation is optimized to 400 Ry. The time constant of the thermostat is set to \SI{100}{ps}, which has shown to be exceptionally good for
preserving vibrational dynamics. The water density is set to \SI{0.998}{g\per cm^3}. The simulation trajectory is 200 ps in length with a timestep of 0.5 fs.
Data are saved in cube files, which present two sets of charge distributions. The electronic charge distribution is presented on 240$^3$ voxels. The voxel size is $l=$\SI{0.0083}{nm}. The nuclei charge distribution consists of  $(x,y,z))$ positions of the 3x256 atoms.

A first processing treatment conserves the delocalized electronic density computed with the AIMD and sums the two electronic and the nuclear charge distribution, into a single charge distribution defined on the voxel grid.
The charge of each nucleus is thus distributed in the 8 closest voxels around the nucleus as follows:
\begin{equation}
	q_i=\frac{\prod_{j=1}^3(l-l_{ij})}{V_0} Q_n
\end{equation}
with $Q_n$, the charge of the nucleus, {\it i.e.} +6$e$ for the oxygen, +1$e$ for the hydrogen, $e$ being the elementary charge. $l_{ij}$ is the distance of the nucleus to the voxel $i$ ($i=1,...,8$) in the direction $j$ ($j=x,y,z$) with $V_0$ the volume of the cube defined by the eight voxels. This linear proportionality rule preserves the dipole moment of the system. The final charge associated with the voxel $i$ is the contribution from the nucleus charge, $q_i$, plus the electronic charge. 
Each simulation frame is thus associated with 240$^3$ charge density values. \par
In a second postprocessing treatment the localized Wannier centers are calculated every \SI{2}{fs}, and assigned to the molecule of the
nearest oxygen, which always results in exactly four Wannier
centers per water molecule. A charge of -2e is assigned to each
Wannier center (see Fig.~\ref{water_molecule}), allowing for the calculation of the dipole
moment. The resulting overall charge distribution is a set of localized charges in the simulation cell. Each simulation frame is now associated with 7 charges per water molecule for 256 molecules, {\it i.e.} 1792 charge density values, which is about 10$^4$ times less than the non-localized charge distribution.

\section{Dielectric response of water}
\subsection{$q=0$ permittivity}
We first compute the macroscopic ($q$=0) $\epsilon$ dielectric permittivity following ~\cite{neuman83}
\begin{align}
\label{permittivity}
	\epsilon -1 = \frac{\left\langle \mathbf{M} \cdot \mathbf{M} \right\rangle - \left\langle \mathbf{M} \right\rangle \cdot \left\langle \mathbf{M} \right\rangle}{3\epsilon_0 k_B T V}
\end{align}
with $\mathbf{M}$ the total dipolar moment of the simulation box. 
The total dipolar moment,
\begin{equation}
	\mathbf{M}=\sum_i \mathbf{m}_i,
\end{equation}
is written as the sum of $\mathbf{m}_i$ the dipole moment of the molecule $i$, with $i \in [1,N_w]$, with $N_w$ the total number of water molecules in the simulation box.
The dipole moment of the molecule $i$ is given by:
\begin{equation}
\label{molecular_dipole}
\mathbf{m}_i=\sum_\alpha q_\alpha \mathbf{r}_{\alpha,i}.
\end{equation}
where we sum over the $\alpha$ charges of a molecule.
We consider the models associated with localised charge distribution, 
 with $\alpha=3$ for SPC/E, SPC/Ef and TIP4P, $\alpha=4$ for TIP5P and $\alpha=7$ for Wannier center treatment, as illustrated in Fig.~\ref{water_molecule}.
The vector  $\mathbf{r}_{\alpha,i}$ denotes the position of the charge $q_\alpha$ to the center of masse of the $i^{th}$ molecule.  For classical models, we compute $\epsilon$ for both short (\SI{200}{ps}) and long (\SI{20}{ns}) trajectories. 
The results are presented in Tab.~\ref{tab:permittivity}.\par
\begin{table}[h!]
	\center
	\begin{tabular}{c|c|c}
	
		model & $\epsilon$ (200 ps) & $\epsilon$ (20 ns) \\
		\hline
		SPC/E & 55.1  & 56.7  \\
		SPC/E/f & 54.4  & 89.8 \\
		TIP4P & 42.5   & 43.5  \\
		TIP5P & 52.3  & 68.0 \\
		Wannier (7 charges) & 48.9  & $\emptyset$ \\

	\end{tabular}
	\caption{Long-range ($q$=0) dielectric permittivity $\epsilon$ for classical and quantum models of water. We compute $\epsilon$ using Eqs~(\ref{permittivity}-\ref{molecular_dipole}) for \SI{200}{ps}  trajectories. For classical MD, we also compute $\epsilon$ for \SI{20}{ns} trajectories.}
	\label{tab:permittivity}
\end{table}

 The dielectric constant is much lower than the expected value of 80 and varies significantly,  between 42 and 55, for short trajectories, depending on the model we consider. For AIMD simulation treated with Wannier centers (last line of Table 1), we find $\epsilon$=49, which is consistently in the same range as classical models. 
 The difference for the $\epsilon$ value between short and long trajectories for classical force-fields is very small for SPC/E, TIP4P, more pronounced for TIP5P, and significant for SPC/Ef. This indicates that temporal relaxation is not fully achieved for the $q=0$ mode for TIP5P and SPC/Ef, in contrast to SPC/E and TIP4P. \par
 
 For long trajectories, $\epsilon$  is lower than those reported in the literature~\cite{elton_these}. This is due to the use of a smaller simulation box than is usually considered in classical MD. We chose this size box for sake of consistency with the DFT data.\par

We attribute the underestimation of $\epsilon$ to the fact that classical models are developed to reproduce experimental values for quantities such as density, and enthalpy of evaporation, but not for the dielectric constant. Note that this caveat has been solved since  a reparametrization of SPC/E has been proposed to reproduce the dielectric
constant of water.~\cite{azcatl2014}\par

\subsection{Nonlocal response function $\chi(q)$}
We are now interested in the wavenumber-dependent (non-local) properties of the system. 
The dielectric response function $\chi(q)$  can be written as a function of the charge structure factor $S(q)$ using the fluctuation-dissipation theorem,
\begin{equation}
\label{response_function}
\chi(q)=\frac{S(q)}{q^2 \epsilon_0 k_BT}.
\end{equation}
For the models associated with localised electronic density, the charge density in Fourier space, $\tilde{\rho}(q)$  can be computed following
\begin{eqnarray}
\label{rhoq}
\tilde{\rho}(q)&=&\sum_{i=1}^{Nw} e^{i \mathbf{q} \cdot \mathbf{r}_i} \sum_{\alpha}q_\alpha e^{i\mathbf{q}\cdot\mathbf{r}_{\alpha,i}},
\end{eqnarray}
with $\mathbf{r}_i$ the position of the center of mass of the molecule $i$.
The charge structure factor follows
\begin{equation}
	S(q)=\frac{\langle \tilde{\rho}(q) \tilde{\rho}(-q) \rangle}{V}\,
\end{equation}
with $V$ the volume of the box. Note that we take into account the periodicity of the system and calculate the charge structure factor for discretized values of the wavenumber $q$. It leads to
\begin{align}
\label{qdiscrete}
q=\frac{2\pi}{L}\sqrt{n_x^2+n_y^2+n_z^2}, \quad {\rm with}  \quad n_x,n_y,n_y \in \mathbb{Z}
\end{align}  
The minimal value of $q$ that can be considered is thus $q=2\pi/L\approx$3 nm$^{-1}$. \par
Now, we compute the charge structure factor associated with the Wannier dipoles extracted from the AIMD simulation. 
First, we calculate the $N_w$ dipoles ${\bf \mu}_i$, with $i\in [1,N_w]$ associated with each water molecule in the simulation box. Then, we  define the response function as follows~\cite{raineri1992,vatin21}
\begin{equation}
\label{wannier_gas}
S(q)=\langle\frac{{\mathbf M({\bf q})}{\mathbf M(-\bf{q})}}{V}\rangle,
\end{equation}
with the total dipole moment ${\mathbf M({\bf q})}$ is equal to:
\begin{equation}
{\mathbf M(q)}=\frac{\bf{q}}{q} \cdot \sum_{i=1}^{N_w}\mu_ie^{i{\mathbf q}\cdot{{\bf r}_i}}.
\end{equation}\par
Finally, we calculate the charge structure factor for AIMD-DFT simulations with the full delocalised electron density.
We  Fourier transform the charge density $\rho(\bf{r})$ defined in real space on the voxel grid  and get $\rho(\bf{q})$ in Fourier space.  \footnote{To do so, we have used the module fast Fourier transform, {\it FFT}, from scipy}.
Then, we use the definition of $\chi(q)$ given in Eq.~(\ref{response_function}).
Note that for this set of data, the wavenumber given in Eq.~(\ref{qdiscrete}) is discretized with  $n_i\in \mathbb{N}, n_i\in[-N/2,N/2]$
where $N=240$ is the number of voxel in each direction (x,y,z). The minimal value of $q$ that we can consider is the same as for the localized charge distribution, but now there is a maximal value for $q$ imposed by the size of the voxel grid, $q_{\rm max}=$650~nm$^{-1}$.

\section{Results and discussion}
\begin{figure*}
	\center
	\includegraphics[scale=0.27]{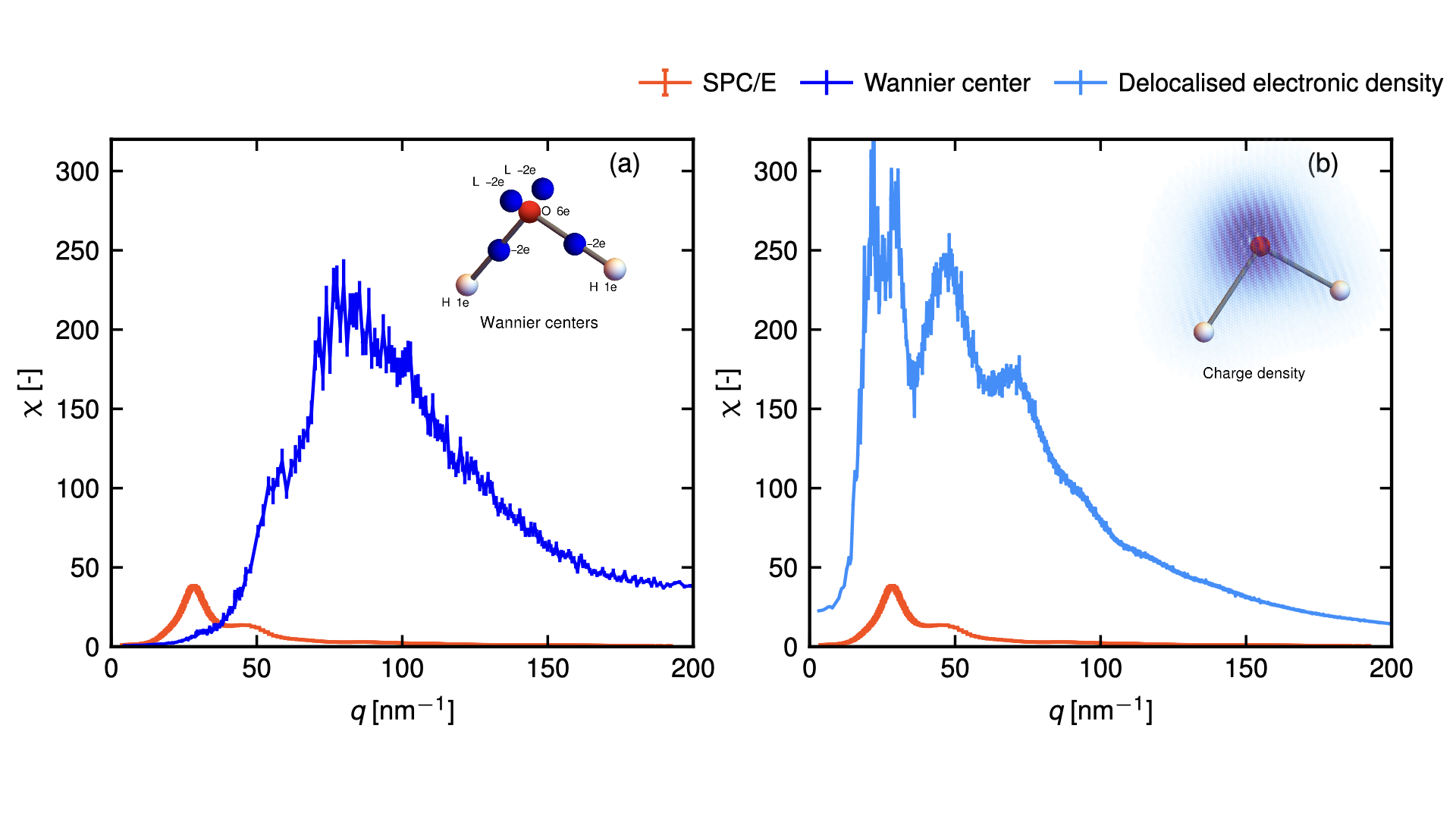}
	\caption{Dielectric response function from AIMD simulations compared with classical MD. (a) $\chi(q)$ for the AIMD-DFT simulations with Wannier center treatment and (b) $\chi(q)$ for the AIMD-DFT simulations with the delocalised electronic density compared to SPC/E dielectric response function. Note that in both cases we check that for large wave numbers the susceptibility decays to the expected value $\chi(q)=0$, as shown in Appendix~\ref{ap:long-range}. Error bars are derived following the reblocking method~\cite{flyvbjerg1989}. $\chi(q)$ for SPC/E has been computed from trajectories of \SI{20}{ns} but the functions already convergees for a trajectory of \SI{200}{ps} (see Appendix~\ref{ap:convergence}), which is the running time of the AIMD DFT simulations.}
	\label{fig2:chi_DFT}
\end{figure*}
We plot the susceptibility $\chi(q)$ obtained from AIMD simulations with a Wannier center post-treatment and with the full delocalised electronic density, and compare it with SPC/E model. The results are presented in Fig.~\ref{fig2:chi_DFT}.
The panel (a) presents $\chi(q)$ for the Wannier center data (blue plot) compared to the one for SPC/E (red plot). For DFT after Wannier projection, we see a unique broad peak around $q\approx$100~nm$^{-1}$ with an amplitude of 200. $\chi(q)$ slowly decreases at larger $q$. The panel (b) shows $\chi(q)$ for the full electronic density (blue plot). 
 The full DFT susceptiblity peak is much more featured than the others and it presents a main double peak for $q$=25~nm$^{-1}$ at an amplitude of $\chi\approx$260 and two satellite peaks, at $q$=45nm$^{-1}$, $q$=60nm$^{-1}$, of amplitude $\chi$=260, $\chi$=200 respectively. \par
We demonstrate here that the non-local dielectric response associated with DFT-based AIMD simulations is completely different from the response obtained with classical MD simulations.
 More surprisingly, considering the Wannier centers instead of the fully delocalised electronic density completely modifies the response.
\subsection{Comparison of 3 rigid models}
To gain further insight into the relationship between the charge distribution assigned to one molecule and the dielectric response function, we decide to look at two other standard classical force fields: TIP4P and TIP5P, and compare them with SPC/E.\par
\begin{figure*}
	\center
	\includegraphics{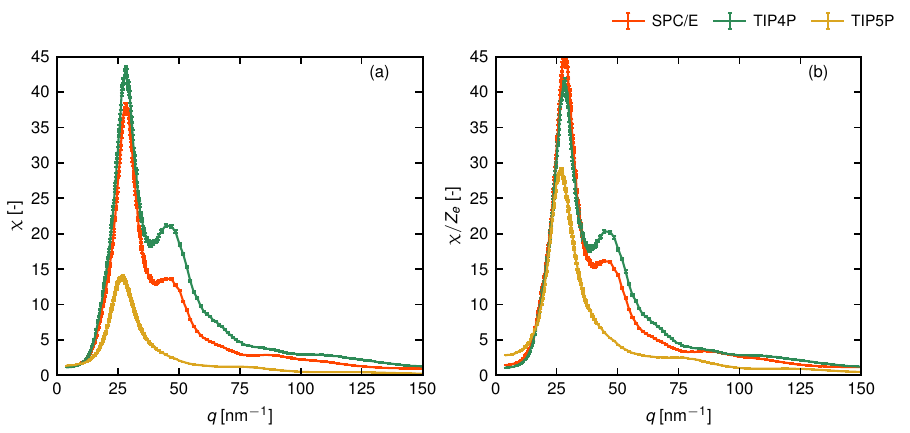}
	\caption{MD simulated response functions $\chi(q)$ for rigid water models in Fourier space. Panel (a) shows the unnormalized result and panel (b) shows $\chi(q)/Z_e$, with $Z_e$ the electronic charge of one molecule. The red curve is obtained for SPC/E ($Z_e=0.841$), the brown curve for TIP4P ($Z_e$=1.04), and the blue curve for TIP5P ($Z_e$=0.482). Response functions are calculated from the long trajectories (\SI{20}{ns}) using the charge structure factor, Eq.~(\ref{response_function}), for discrete wavenumbers with a minimal $q=2\pi/L\approx$30nm$^{-1}$ fixed by the size $L$ of the simulation box.}
	\label{classicalmodels}
\end{figure*}

The panel (a) of Fig.~\ref{classicalmodels} shows the response function $\chi(q)$ for SPC/E (red plot), TIP4P (brown plot), TIP5P (pink plot).
The two three-point charge models SPC/E and TIP4P present very similar curves, with a main peak for $q\approx$~30~nm$^{-1}$, with an amplitude of about 40, and a satellite peak (around $q\approx$~50~nm$^{-1}$.) TIP5P is characterized by a single peak for $q\approx$~26~nm$^{-1}$, less sharp and with an amplitude divided by 3 when compared to the 2 other models. 
We recall here that the response function $\chi$ gives the charge induced in water $\rho_{\rm ind}$ by an external charge, $\rho_{\rm ext}$, typically an ion. In Fourier space, the relationship between $\rho_{\rm ind}$ and $\rho_{\rm ext}$  can be simply written as:
\begin{equation}
 \rho_{\rm ind}(q)=-\chi(q)\rho_{\rm ext}(q).
\end{equation}
For rigid classical models, the dielectric response of the fluid is mainly due to the orientation of the molecules. The resulting induced charge is proportional to the amount of charge present in a molecule of water and this molecular charge varies a lot with all the models for water. We rescale $\chi(q)$ by the amount of electronic charge $Z_e$ in the water molecule, which is $Z_e$=0.847 for SPC/E,$Z_e$=1.04 for TIP4P and $Z_e$=0.482 for TIP5P. The panel (b) of Fig.~\ref{classicalmodels} shows $\chi(q)/Z_e$ for the 3 classical force fields. The three plots show now a more similar structure, the amplitude of the overresponse peak varying between 30 for TIP5P and 45 for SPC/E. This confirms our assumption above  that the amplitude of the response encodes the total amount of the electronic charge on one molecule of the water model. \par
In contrast, the {\it ab initio} model for water includes all the valence electrons of the molecule and its electronic charge number, $Z_e$=8, is much larger than that of the classical models. This explains the large difference in the amplitude of $\chi(q)$ seen in Fig.~\ref{fig2:chi_DFT}.
\subsection{AIMD simulated versus SPC/E simulated susceptibility}
\begin{figure}
	\center
	\includegraphics[]{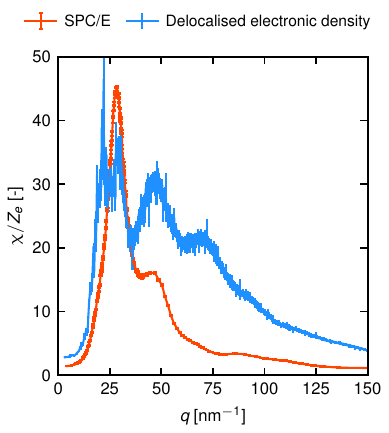}
	\caption{Rescaled susceptibility for AIMD and SPC/E simulations. $Z_e$=8 for AIMD simulations and $Z_e$=0.841 for SPC/E. The response function for SPC/E is obtained from long trajectories of \SI{20}{ns}.}
	\label{chi_renormalisedDFTSPC/E}
\end{figure}

With this in mind, we plot the renormalised susceptibility of the AIMD simulation with delocalised electronic density and compare it with the SPC/E renormalised susceptibility.
Figure~\ref{chi_renormalisedDFTSPC/E} shows  $\chi(q)/Z_e$ for  SPC/E (red plot) and the AIMD simulated $\chi(q)/Z_e$, with delocalised electronic density (blue plot). We can now compare the two dielectric response functions as the amplitudes are comparable and the two curves overlap well. First we consider the intermediate range of $q$ values $q \in [10-50]$~nm$^{-1}$, corresponding to distances $\lambda=2\pi/q$ of a few nm, corresponding to intermolecular distances. Even at this scale, where classical force fields are supposed to be reliable, we can see that the full DFT simulated response function contains a structure at the lowest $q$ (slightly below 25~nm$^{-1}$) that is missing in the SPC/E model.  We suggest that the main peak is split into two components and slightly shifted to the low $q$-regions.  The splitting on the main peak was also obtained by retrospectively including a delocalisation of the charges to the classical force-field charge structure factor~\cite{kornyshev1999}. This suggests two characteristic lengths to structure the water H-bonding network, whereas the SPC/E model highlights a single characteristic length. At larger $q$, $q\geq$50~nm$^{-1}$, the full DFT signal shows two extra peaks, one around $q$=50~nm$^{-1}$ and a smaller peak around $q$=75~nm$^{-1}$. The peak at 50 is present in the SPC/E signal albeit much reduced in amplitude. The highest-q peak at 75 seems to emerge from electronic charge correlation in the sub-angstroms length scale and therfore is absent for the SPC/E signal.~\cite{dufils2024}.
\subsection{Dielectric response for the Wannier centers approach}
We are now back to the response function that was obtained for the AIMD simulations post-treated with Wannier center approach. As shown in Fig.~\ref{fig2:chi_DFT} panel (a), the correlations obtained for the charge distribution composed of Wannier centers and nuclei of water do not make sense. $\chi(q)$ at low $q$ completely misses the main peak also called the "overreponse" peak. 
To unveil the origin of the wrong Wannier response, we treat the AIMD data as follows. First, we keep only the DFT atomic positions of the 3 nuclei to which we assign the charges of the SPC/E model, see (DFT-3 charges) in Fig.~\ref{Wannier}. We then consider 5 centers: the 3 nuclei constitutive of water and the 2 Wannier centers corresponding to the lone pairs of the oxygen of the water molecule and assign the charges according to the TIP5P model, as given in Fig.~\ref{water_molecule}. See DFT-5 charges in Fig.~\ref{water_molecule}. Doing so, we recover the "overresponse peak", obtain good agreement with the corresponding classical force-fields. The results are presented in Appendix~\ref{ap:wannier-centers}. This confirms to us that the inconsistent response function using the Wannier post-treatment including 7 charges is due to the very high values of localised charges that appear in the Wannier center treatment. \par
\begin{figure}
	\center
	\includegraphics[]{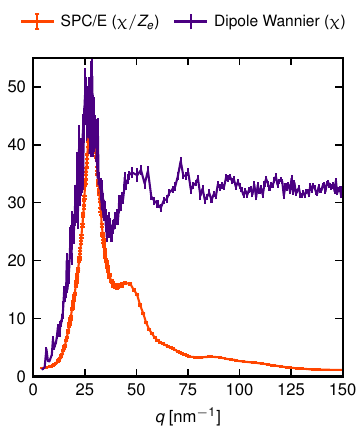}
	\caption{Susceptibility for SPC/E water model and Wannier dipoles from AIMD simulations. We plot the rescaled susceptibility for SPC/E (orange plot) that we compare to the susceptibility of water dipoles computed from the Wannier centers (purple plot).}
	\label{chi_dipoleWC}
\end{figure}
Next, we consider the dipole associated to each water molecule from the Wannier centers and the nuclei.  We compute the dielectric response function of this liquid of dipoles, as detailed in Eq.~(\ref{wannier_gas}). The figure~\ref{chi_dipoleWC} shows the resulting $\chi(q)$ compared to the SPC/E one. The amplitude and the location of the "over-response" peak  around $q$=30~nm$^{-1}$ are in a good agreement: the dipole-derived from Wannier centers do reproduce correctly reproduces the dielectric properties of SPC/E water validating the use of this methodology for describing the dipole function. At larger wavenumbers, $q\geq$40~nm$^{-1}$, the signal oscillates around an unphysical nonvanishing value instead of decaying to zero as expected. This comes from the charge structure factor, given in Eq.~(\ref{wannier_gas}) for this case. The fluid is treated as a distribution of point dipoles. At short length scales, comparable to the charge separation distance in water, i.e. at large $q$ values, this reduction leads to inadequate results because the neglected multipole contributions become increasingly important.~\cite{raineri1992}.

\subsection{Dielectric response functions of more sophisticated water models}
\begin{figure}
	\center
	\includegraphics[]{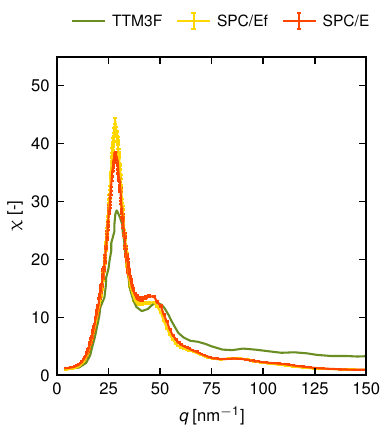}
	\caption{Suscpetibility for SPC/E, SPC/Ef and TTM3F water models.  We plot $\chi(q)$ for SPC/E and SPC/Ef and reproduce the susceptibility of TTM3F polarisable and flexible model from ref.~\cite{elton_these}.}
	\label{polandflex}
\end{figure}

Now that we have illuminated the properties of the AIMD simulated response function compared to the rigid classical force fields of water, we wonder if more sophisticated classical water models could give closer properties to the quantum response at low computational cost.
We first consider a flexible SPC/E model, illustrated in Fig.~\ref{water_molecule} and Eq.~(\ref{k_flex}) and compute its associated response function, which is represented in orange on Fig.~\ref*{polandflex}.
The only effect of the flexibility is a small increase in the amplitude of the main peak. It indicates a strengthening of the H-bond network structuring the fluid.  
The response function is not modified for large wavenumbers $q \geq $ 50~nm$^{-1}$. The flexibility does not mimic the DFT electronic response that we have seen appear at large $q$.
For a polarizable model, we reproduce published data from the TTM3F model~\cite{fanourgakis} from the ref.~\cite{elton_these}. This model is a four-site model, similar to TIP4P, which is also flexible. It contains one polarization dipole per molecule located on the extra M-site. It also contains fluctuating charges, but their effect on polarization has been measured to be weak~\cite{elton2014}. The response function is plotted in dark green on Fig.~\ref{polandflex}.
The lower magnitude for TTM3F is because the model has about 50$\%$ less charge on each atom compared to SPC/E model. Similar to SPC/E, it shows a main peak at $q$=30~nm$^{-1}$ and a satellite peak at  $q$=50~nm$^{-1}$.  $\chi(q)$ reaches an uniphysical limit at large $q$, due to the inclusion of punctual dipoles, as already commented for Wannier dipoles, see Fig.~\ref{chi_dipoleWC}, and thus cannot well describe large $q$ structure.

We conclude that neither flexible nor polarisable classical models describe nonlocal properties at large $q$ which are comparable to DFT models.

\section{Conclusion}
In this work, we study the nonlocal dielectric properties of liquid water within classical force-field models and beyond at the electronic scale using DFT. We compute the dielectric response function of water, including the delocalised electronic charge density, using AIMD simulations. We show that the result is very different from those obtained using classical force-fields, whether rigid, flexible or polarisable. 
The amplitude of the dielectric response of DFT water is much larger than the classical one, due to the higher charge density in the system. Also, the range of wavenumbers over which the system responds is increased  toward high-$q$ values. AIMD simulations identify more typical lengths that characterise the charge distribution in the liquid, especially at small distances characteristic of  intramolecular lengths where electronic density matters.\par

We are aware that AIMD simulations, while more accurate, are expensive in terms of computational resources. This drawback has recently stimulated the development of neural network quantum molecular dynamics to combine the speed of classical methods with the precision of quantum simulations.
Recently, deep neural networks trained on Wannier center positions~\cite{omranpour24,gao2024} or other electronic
observables such as Wannier displacements~\cite{jana2024} have
been developed with the aim of accurately describing
long-range electrostatic interactions and electronic. As
noted here, the charge distribution associated with localised
Wannier centers leads to an unphysical dielectric
response function, so one could envisage training the networks
with delocalised functions.\par
The dielectric response of water at the electronic scale is at the heart of current questions, from the 'quantum friction' between water and graphene~\cite{kavoline2022} to the rate of electron transfer reaction~\cite{jeanmairet2018}. Simulations addressing these issues include the electronic degrees of freedom for the metal but usually retain the classical description of the liquid. We show here that this can lead to a large misestimation of the studied effect since the dielectric properties of the AIMD-DFT and the classical force-field water model are not superimposed at all.

\section*{Acknowledgements}
HB and DL thank Anthony Maggs for help with data processing. HB and DL thank CNRS for EMERGENCE@INC2023 funding.

\renewcommand\thesection{}  

\section*{Appendix}
\renewcommand\thefigure{A.\arabic{figure}}   
\setcounter{figure}{0} 
\subsection{Large wavenumber limit}\label{ap:long-range}
\begin{figure*}
	\center
	\includegraphics{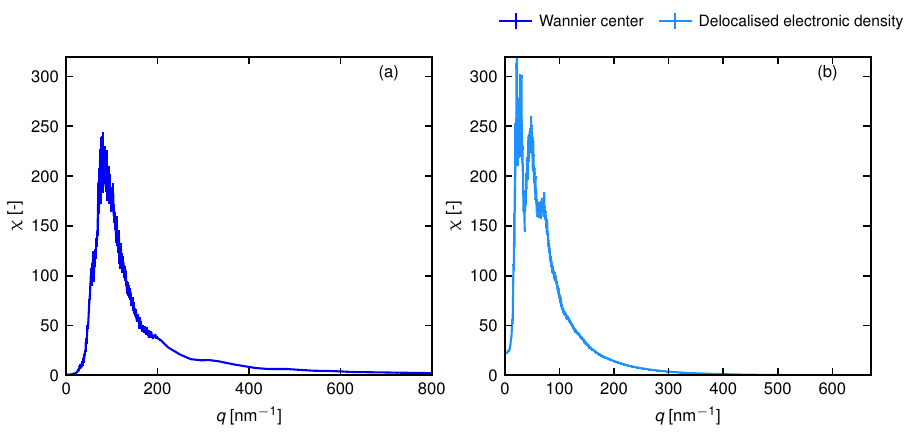}
	\caption{Dielectric response from AIMD simulations for (a) electronic charges located on the Wannier centers, (b) full delocalised electronic density. }
	\label{figA:largeq}
\end{figure*}

\begin{figure}
	\center
	\includegraphics{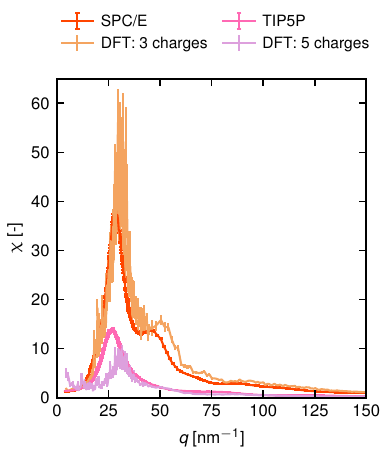}
	\caption{Dielectric response from AIMD-DFT and classical
simulations. The light orange curve corresponds to the re-
sponse function of a 3 charge model constructed from AIMD-DFT simulations and to the SPC/E model. The light purple curve corresponds to the response function of a 5 charge model
constructed from AIMD-DFT simulations and to the TIP5P
model.}
	\label{Wannier}
\end{figure}

Here, in Fig.~\ref{figA:largeq} we check the limit of $\chi(q)$ at large $q$ for AIMD simulation, for both for the use of Wannier center distribution (left panel) and for the full delocalised electronic charge (right panel). As expected, we see that $\chi(q)$ tends to zero.
\subsection{Dielectric response function of DFT nuclei carrying classical force-field charges}\label{ap:wannier-centers}
Here, we compute the dielectric response function for
a 3 point-charge model and a 5 point-charge model constructed
from AIMD-DFT data and compare it to classical force-fields results. First, we keep only the DFT atomic positions of the 3 nuclei constitutive of water to which we assign the charges of the SPC/E model (light orange plot). We then consider 5 centers: the 3 nuclei and
the 2 Wannier centers corresponding to the lone pairs of the oxygen of the water molecule and assign the charges according to the TIP5P model (light purple plot). We
compare these results for SPC/E (red plot) and TIP5P (pink plot) respectively. We see a good agreement between the DFT and classical point-charge models.
\subsection{Convergence of the response function for classical models after \SI{200}{ps}}\label{ap:convergence}
In this section, we compare the response function $\chi(q)$ for the flexible version of SPC/E for short runs of \SI{200}{ps} and long runs of \SI{20}{ns}. The grey curve corresponds to the analysis of the short trajectory, the orange curve corresponds to the long one. As can be seen, the convergence is already very good after \SI{200}{ps}. This shows that there is no problem in comparing DFT data from short trajectories with  classical models data from long trajectories.
\begin{figure}
	\center
	\includegraphics{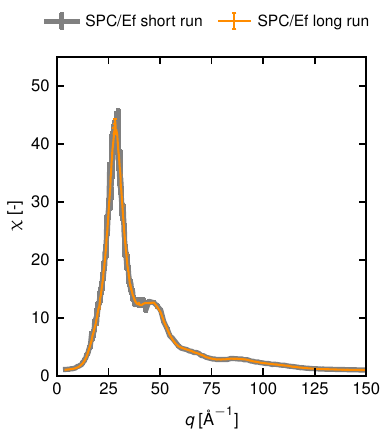}
	\caption{Dielectric response function for SPC/E flexible model
for short trajectory 200 ps (grey curve) and long trajectory
20 ns (orange curve).}
	\label{figA1:converg}
\end{figure}

\bibliographystyle{unsrt}
\bibliography{RPfinal}
\end{document}